\documentclass[
longbibliography,
preprint,
pre,
amsmath,amssymb,
aps
]{revtex4-2}
\usepackage[T1]{fontenc}
\usepackage{graphicx}
\usepackage{dcolumn}
\usepackage{bm}

\usepackage{amsmath}
\allowdisplaybreaks[4]
\usepackage{amssymb}
\usepackage{xcolor}
\usepackage{tikz}
\usepackage{placeins}
\begin{document}

\title{When Heterogeneity Drives Hysteresis: Anticonformity in the Multistate $q$-Voter Model on Networks}

\author{Arkadiusz Lipiecki}
    \email{arkadiusz.lipiecki@pwr.edu.pl}
\author{Katarzyna Sznajd-Weron}
 \email{katarzyna.weron@pwr.edu.pl}
\affiliation{
 Department of Computational Social Science\\Wrocław University of Science and Technology, Poland
}
\date{\today}
\begin{abstract}

Discontinuous phase transitions are closely linked to tipping points, critical mass effects, and hysteresis, phenomena that have been confirmed empirically and recognized as highly important in social systems. The multistate $q$-voter model, an agent-based approach to simulate discrete decision-making and opinion dynamics, is particularly relevant in this context. Previous studies of the $q$-voter model with anticonformity on complete graphs uncovered a counterintuitive result. Changing the model formulation from the annealed (homogeneous agents with varying behavior) to quenched (heterogeneous agents with fixed behavior) produces discontinuous phase transitions. This is contrary to the common expectation that quenched heterogeneity smooths transitions. To test whether this effect is merely a mean-field artifact, we extend the analysis to random graphs. Using pair approximation and Monte Carlo simulations, we show that the phenomenon persists beyond the complete graph, specifically on random graphs and Barabási-Albert scale-free networks. The novelty of our work is twofold: (i) we demonstrate for the first time that replacing the annealed with the quenched approach can change the type of phase transitions from continuous to discontinuous not only on complete graphs but also on sparser networks, and (ii) we provide pair-approximation results for the multistate $q$-voter model with competing conformity and anticonformity mechanisms, covering both quenched and annealed cases, which had previously been studied only in binary models.
\end{abstract}

\maketitle
\section{Introduction}
Discontinuous phase transitions in models of opinion dynamics are of particular interest \cite{Li2016DiscontinuousModel,Vieira2016PhaseNoises,Chen2017First-orderInertia,Encinas2018FundamentalModel,Oestereich2020HysteresisNetworks,Malarz2023ThermalGraphs,Krawiecki2023Q-neighborNodes,Krawiecki2024Q-voterApproximations,Chmiel2024Q-NeighborNetwork} because they are often associated with tipping points, critical mass phenomena, and social hysteresis \cite{Sznajd-Weron2024TowardModeling}. The $q$-voter model \cite{Castellano2009NonlinearModel}, which can be also interpreted as the binary-choice dynamics \cite{DalForno2018ReferenceDynamics}, provides a simple yet powerful framework to study such effects. When competing mechanisms, such as conformity and independence, or conformity and anticonformity (known also as the contrarian behavior \cite{Galam2004ContrarianScenario}), interact, they can drive phase transitions between consensus, polarization, or disordered states. One of the most important factors influencing these transitions is how the mechanisms are implemented, either through an annealed approach, where individual preferences change over time, or a quenched approach, where preferences remain fixed \cite{Jedrzejewski2025WhenDynamics}.

In the binary $q$-voter model with anticonformity, only continuous phase transitions have been observed for both the annealed and quenched approaches. While pair approximation (PA) suggested the possibility of a discontinuous transition, computer simulations demonstrated that this change does not occur, indicating that it is merely an artifact of the PA method \cite{Jedrzejewski2022PairNetworks}. In contrast, in the $q$-voter model with independence, the quenched approach either eliminates the discontinuous transition in the binary case \cite{Jedrzejewski2022PairNetworks} or smooths it in multistate versions \cite{Nowak2021DiscontinuousDisorder}, whereas such transitions are observed under annealed dynamics. This behavior is consistent with well-established results from statistical physics, where quenched heterogeneity typically weakens or destroys discontinuous phase transitions \cite{Imry1975Random-fieldSymmetry,Aizenman1989RoundingDisorder,Binder1987TheoryTransitions,Munoz2010GriffithsNetworks,Odor2021HeterogeneousTransitions}.

The unexpected behavior appears when the $q$-voter model with anticonformity is generalized to multiple states. In this case, it has recently been shown that switching from annealed to quenched disorder does not eliminate discontinuities, but instead induces them: in the annealed approach the transition is continuous, whereas under the quenched approach it becomes discontinuous \cite{Nowak2022SwitchingDisorder}. This inversion of the usual quenched-annealed relationship poses a conceptual puzzle, as it contradicts established expectations from statistical physics.

One natural concern is that this effect might be an artifact of the mean-field approximation, since the previous study was restricted to complete graphs. To address this, in the present work, we extend the analysis of the multistate $q$-voter model with anticonformity beyond complete graphs. Using PA and Monte Carlo simulations, we investigate the model on random, random regular, and Barabási–Albert scale-free networks~\cite{Barabasi1999}. 

Our results show that the puzzling behavior persists beyond complete graphs: while the annealed approach consistently produces continuous transitions, the quenched approach, for the $q$-voter model with more than two states, yields discontinuous transitions accompanied by hysteresis. Thus, this phenomenon is not a mean-field artifact, but a genuine effect of heterogeneity introduced by quenched anticonformity. 

The novelty of this paper is two-fold. First, to the best of our knowledge, this is the first study that shows that replacing the annealed with the quenched approach can change the type of phase transition on random graphs from continuous to discontinuous. Second, for the first time, we present PA results for the multistate $q$-voter model with competing mechanisms, specifically conformity and anticonformity. Moreover, since these competing mechanisms can be implemented in both quenched and annealed forms, we provide PA results for both cases. Previously, such analyses were performed only for binary models \cite{Jedrzejewski2017PairNetworks,Gradowski2020PairNetworks,Jedrzejewski2022PairNetworks}.

\section{Model}
We consider a system of $N$ agents, also called voters, positioned at the vertices (nodes) of an arbitrary graph with $N$ vertices. Each vertex $v \in \{1, \ldots, N\}$ is occupied by exactly one voter, and each voter is assigned a dynamical variable $s_v(t)$ representing its state, which can take one of $S$ possible values:
\begin{equation}
s_v(t) = s, \quad s \in \{1, \ldots, S\}.
\end{equation}
This state can be interpreted in various ways (opinion, belief, attitude, etc.) \cite{Olsson2024AnalogiesDynamics}. In this model, we specifically treat it as a categorical choice among $S$ alternatives, corresponding to discrete choices in which a decision-maker selects one option from a finite set \cite{Fosgerau2018AMaximum}. Although the alternatives are labeled $1, \ldots, S$, this labeling is purely for enumeration; there is no ordering or 'distance' between them, unlike Likert scales or models where neighboring states matter \cite{Lipiecki2025DepolarizingAnticonformity}. 

We define agents as neighbors when they occupy vertices that are connected by an edge (link). The opinion of a target (focal) voter can change when influenced by a unanimous group of $q$ neighbors. If, within such a group, the opinion of at least one agent differs from the others, then the group does not exert any influence on the target. As in some other versions of the $q$-voter model, the influence group, also called the $q$-panel, the source of influence, or simply a source, is formed by drawing a group of $q$ agents from the neighborhood of the target voter without repetition \cite{Nyczka2012PhaseDriving,Jedrzejewski2017PairNetworks,Jedrzejewski2022PairNetworks,Nowak2022SwitchingDisorder,Mullick2025SocialInfluence} . Note that in the original $q$-voter model \cite{Castellano2009NonlinearModel}, and many later modifications of the model \cite{Mobilia2015NonlinearZealots,Mellor2017HeterogeneousZealotry,Vieira2018ThresholdModel,Vieira2020PairModel,Ramirez2024OrderingModels}, repetitions are allowed.

Following \cite{Nowak2022SwitchingDisorder}, we consider two types of behaviors, describing the response of a voter to the influence of the $q$-panel: conformity and anticonformity. If a target voter acts as a conformist, it adopts the state of a unanimous $q$-panel, while an anticonformist flips to a randomly selected state (other than its current state) provided that all the voters in the $q$-panel are in the same state as the target.

Within an \textit{annealed approach} also referred to as \textit{annealed disorder}, the behavior of each voter is independent at each time step: whether a voter acts as an anticonformist is a Bernoulli trial with success probability $p$, and otherwise the voter acts as a conformist with complementary probability $(1-p)$. Hence, in the annealed approach, all voters are homogeneous with respect to behavior: every voter is equally likely to act as a conformist or an anticonformist. In the case of the \textit{quenched approach}, also referred to as \textit{quenched disorder}, the behavior of each voter is described by a static variable (trait):
\begin{equation}
\eta_v = \eta, \quad \eta \in \{\bullet,\circ\} \equiv \{\mbox{anticonformist,conformist}\}.   
\end{equation}
It is still a Bernoulli random variable with probability $p$, but it is drawn only once in the initial state and then remains fixed over time. Therefore, the expected number of anticonformists is $pN$, and the expected number of conformists is $(1-p)N$.

\begin{figure}
    \centering
    \begin{minipage}{0.49\linewidth}
        \centering
        \textbf{Annealed disorder} \\[0.5em]
        \includegraphics[width=\linewidth]{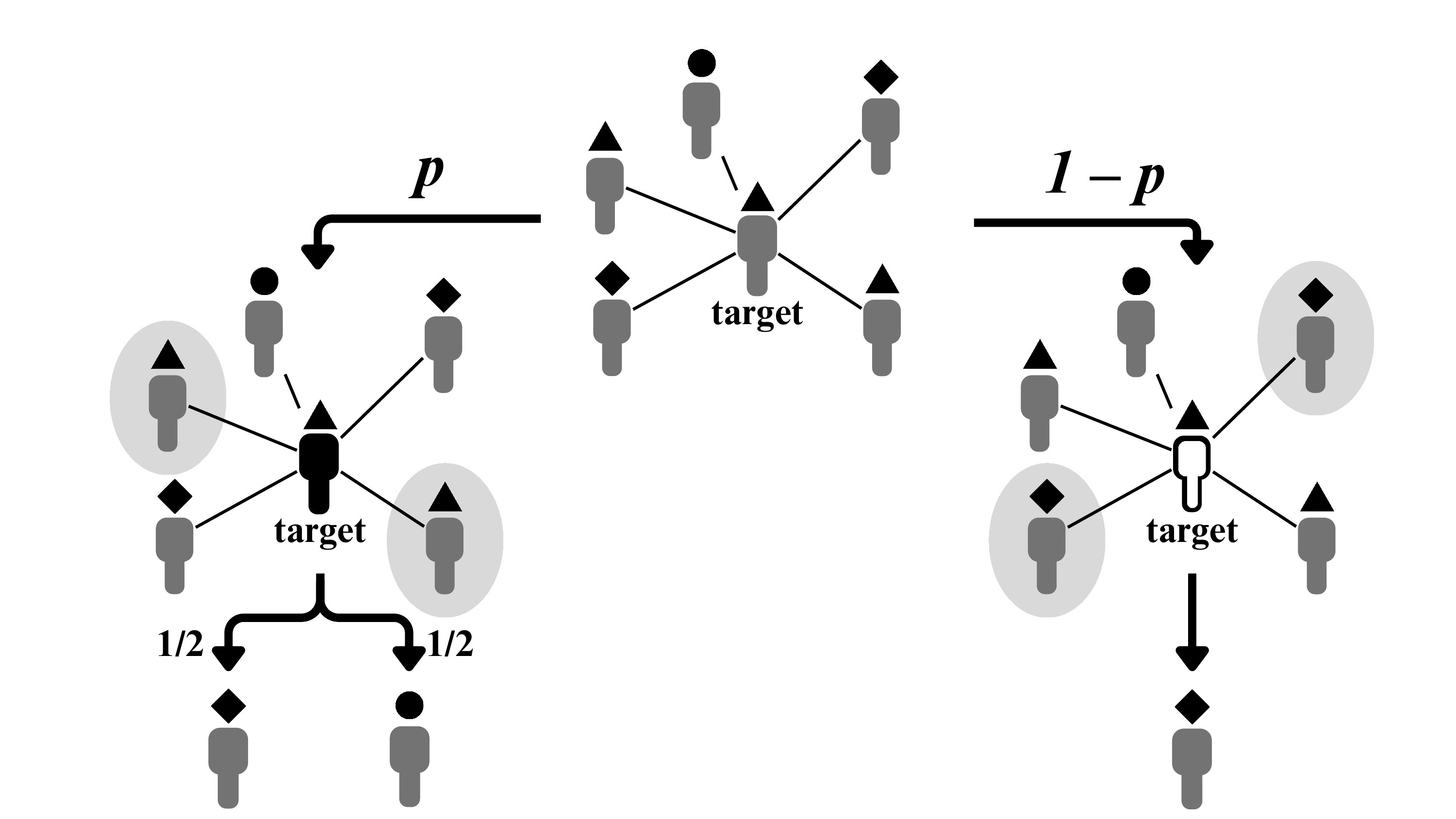}
    \end{minipage}
    \begin{minipage}{0.49\linewidth}
        \centering
        \textbf{Quenched disorder} \\[0.5em]
        \includegraphics[width=\linewidth]{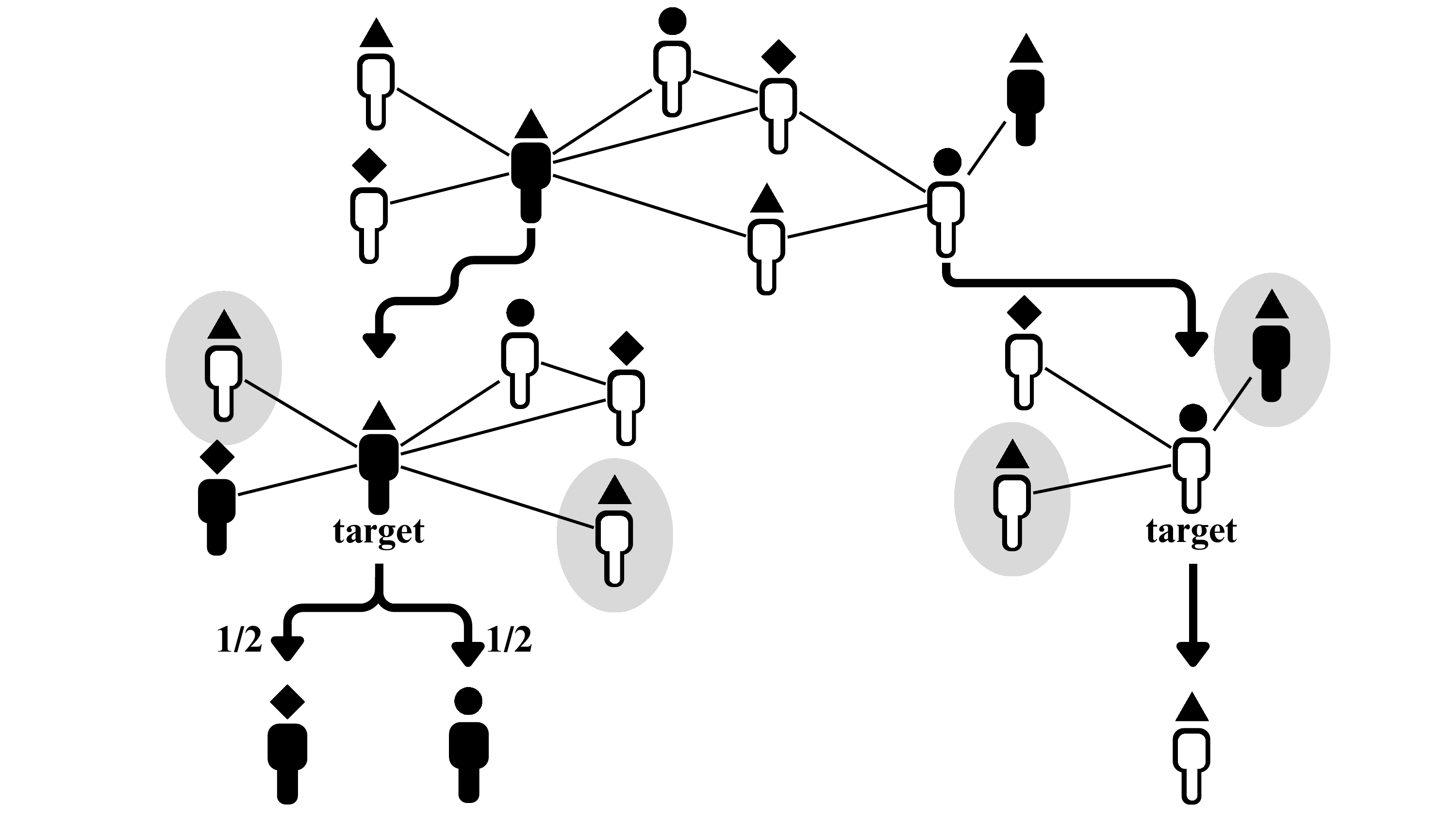}
    \end{minipage}
    \caption{Symbolic scheme of the updating procedure in the annealed (left panel) and quenched (right panel) three-state $q$-voter model with anticonformity, where \scalebox{1.1}{$\blacktriangle$}, \scalebox{1.4}{$\bullet$}, and \scalebox{0.75}{\rotatebox{45}{$\blacksquare$}} denote different states, black-filled voters mark anticonformists, white-filled voters mark conformists. In the annealed version the voters are gray to highlight that their behavior is randomly decided during an update. The neighbors chosen for the $q$-panel are placed inside gray ellipses. In this example, $q=2$.}
    \label{fig:scheme}
\end{figure}

As usual, we use random sequential updating. A unit of time ($t \rightarrow t + 1$) is defined as $N$ elementary updates, which corresponds to one Monte Carlo step (MCS). An elementary update (schematically shown in Fig.~\ref{fig:scheme}) consists of the following steps:  
\begin{enumerate}
    \item Randomly select a target voter $v$.
    \item Randomly select a group of $q$ neighbors of $v$, without repetitions (the $q$-panel).
    \item Check whether all $q$ neighbors are in the same state, i.e., whether the $q$-panel is unanimous. If the $q$-panel is unanimous, proceed to step 4; otherwise, nothing happens.
    \item Update the state of the target voter $v$ according to the version of the model:
    \begin{enumerate}
        \item \textbf{Annealed}: with probability $p$, the target voter $v$ adopts randomly one of the $S-1$ states different from that of the $q$-panel, if the opinion of the $q$-panel is the same as the opinion of the target (anticonformity), and with complementary probability $1-p$ it adopts the same state as the $q$-panel (conformity).  
        \item \textbf{Quenched}: if the target voter $v$ is an anticonformist, it adopts randomly one of the $S-1$ states different from that of the $q$-panel provided that the opinion of the $q$-panel is the same as the opinion of the target; if it is a conformist, it adopts the same state as the $q$-panel.  
    \end{enumerate}
\end{enumerate}

\section{Pair Approximation}
In previous work, the mean-field approximation (MFA) was applied to the multistate $q$-voter model with anticonformity, both in the annealed and quenched formulation, and compared with simulations on a complete graph \cite{Nowak2021DiscontinuousDisorder}. Although MFA is relatively simple to derive and solve, it often becomes inaccurate beyond the complete graph because it ignores correlations between nodes. Here we will use PA, which is a mean-field-like method that improves upon MFA by incorporating dynamical correlations at the pairwise level \cite{Oliveira1993NonequilibriumBehaviour}.

PA has been successfully applied to various binary-state \cite{GleesonBinary-StateBeyond} and multistate dynamics \cite{Fennell2019}, including the voter model \cite{Vazquez2008AnalyticalNetworks}, several versions of the binary $q$-voter model \cite{Jedrzejewski2017PairNetworks,Vieira2020PairModel,Gradowski2020PairNetworks,Jedrzejewski2022PairNetworks,Krawiecki2024Q-voterApproximations}, and more recently to a multistate $q$-voter model \cite{Ramirez2024OrderingModels}. In the latter work, agents could only conform to their neighbors and the focus was on ordering dynamics. The model studied in \cite{Ramirez2024OrderingModels} does not exhibit phase transitions, as there is no competition between different social response mechanisms. Furthermore, all agents are, by definition, identical in their responses to social influence, since only one type of response is allowed. Consequently, the distinction between quenched and annealed heterogeneity of responses is not applicable in this context.

In this paper, for the first time, we obtain PA results for the multistate $q$-voter model with competing mechanisms. Furthermore, since these competing mechanisms can be introduced in both quenched and annealed forms, we present PA results for both cases. Noteworthy, Fennel \& Gleeson \cite{Fennell2019} introduced generalized approximation frameworks for multistate dynamical processes, which encompass the $q$-voter model considered in this study. However, they apply the heterogeneous PA~\cite{Pugliese2009}, hence the number of of independent variables in their general approach scales linearly with the range of possible node degree values. In our application we adopt the homogeneous PA~\cite{Jedrzejewski2017PairNetworks,Jedrzejewski2022PairNetworks, Ramirez2024OrderingModels}, in which the probability of drawing a neighbor in a given state is independent of node degree. This leads to the closed set of evolution equations that scales in size only with the number of states. Furthermore, we treat the multinomial distribution of opinions of neighboring agents~\cite{Fennell2019, Ramirez2024OrderingModels} as a chain of conditional binomial distributions, which allows to simplify the final equations for the $q$-voter dynamics.

The overarching goal of approximating the dynamics of the $q$-voter model is to obtain an analytically or, at least numerically traceable evolution equations for the concentration of voter classes, where the \textit{class} of a voter $v$ represents its state $s_v(t)$ and type $\eta_v$. The concentration of voters in class $(s, \eta)$ is defined as
\begin{eqnarray}
    c_{s,\eta}(t) := \frac{|v\in\{1, ..., N\}: s_v(t)=s \wedge \eta_v = \eta|}{N}.
\end{eqnarray}
In the case of the quenched disorder the type $\eta$ corresponds to behavior, i.e. \textit{conformist} $\circ$ or \textit{anticonformist} $\bullet$, while for the annealed disorder the type is omitted and the class of a voter is only described by its state:
\begin{eqnarray}
    c_{s}(t) := \frac{|v\in\{1, ..., N\}: s_v(t)=s|}{N}.
\end{eqnarray}
In the remainder of the paper, we will use $s$ and $\sigma$ to denote state values, while $\eta$ and $\beta$ will correspond to type values.
Both $c_{s}(t)$ and $c_{s,\eta}(t)$ are, in principle, random variables. However, in the $N \rightarrow \infty$ limit, they converge to their expected values. Therefore, for large systems, we can treat $c_{s}(t)$ and $c_{s,\eta}(t)$ as deterministic variables, which evolution is given by ordinary differential equations (ODEs). Thus, we can describe the time evolution of voter class concentrations with the following equation:
\begin{eqnarray}
    \frac{dc_{s, \eta}}{dt} = \sum_{s'} c_{s', \eta}(t)f_\eta^{s' \rightarrow s} - c_{s, \eta}(t)f_\eta^{s \rightarrow s'},
\label{eqn:general}
\end{eqnarray}
where $f_\eta^{s \rightarrow s'}$ is the probability that a voter in state $s$ and type $\eta$ flips its state to $s'$. The coarsest approximation is based on the assumption that the state concentrations in the neighborhood of each voter are equivalent to the global ones, which corresponds to the mean-field approximation. Within MFA, the system of voters can be fully described by tracking only the concentration of each voter class. The mean-field flipping probabilities for the annealed and quenched disorders are, respectively, given by:
\begin{eqnarray}
    f^{s \rightarrow s'} &=& (1-p)c_{s'}^q + \frac{p}{S-1}c_s^q, \\
    f_\eta^{s \rightarrow s'} &=& \delta_{\eta\circ}c_{s'}^q + \frac{\delta_{\eta\bullet}}{S-1}c_s^q,
\end{eqnarray}
where $\delta_{ij}$ is the standard Kronecker delta.
However, as stated at the beginning of this section, the applicability of the mean-field approximation is very limited, since its assumptions correspond to all-to-all interactions, either in the form of a complete graph topology for the simplest homogeneous approach or an annealed network approximation for the heterogeneous mean-field~\cite{Sood2008, Moretti2012, Moretti2013}. Therefore, in this paper, we use the pair approximation, which accounts for the dynamical correlation between nearest neighbors, allowing us to study the $q$-voter dynamics on sparse networks with negligible clustering.

Within PA, we assume that the state of each neighbor of a given voter $v$ is an independent and identically distributed (i.i.d.) random variable, conditioned only on the class of voter $v$. Hence, the flipping probabilities for the annealed and quenched disorders, respectively, can be expressed in the PA regime as:

\begin{eqnarray}
    f^{s \rightarrow s'} &=& (1-p)(P\left[s'|s\right])^q + \frac{p}{S-1}(P\left[s|s\right])^q, \label{eqn:general:pa:annealed}\\
    f_\eta^{s \rightarrow s'} &=& \delta_{\eta\circ}(P\left[s'|(s, \eta)\right])^q + \frac{\delta_{\eta\bullet}}{S-1}(P\left[s|(s, \eta)\right])^q,
    \label{eqn:general:pa:quenched}
\end{eqnarray}

where $P\left[s|s'\right]$ and $P\left[s|(s', \eta)\right]$ are the probabilities of drawing a neighbor in state $s$ given that the class of the focal voter is $s$ in the annealed approach and $(s, \eta)$ in the quenched approach. For the sake of calculations, we employ the notion of directed edges, artificially replacing each undirected edge in the given graph with two oppositely directed edges, following the convention introduced in~\cite{Jedrzejewski2022PairNetworks}. Thus, the process of randomly selecting a neighbor in class $(s', \eta')$ of the focal voter is equivalent to drawing an out-edge starting at the focal voter's vertex and ending at a vertex with a voter in class $(s', \eta')$. By assuming that such draws are i.i.d. and conditional only on the class of the focal voter, we approximate their probabilities using the concentration of edges linking different voter classes.

\subsection{Annealed disorder}
\begin{figure}[htbp]
    \centering
    \begin{tikzpicture}[scale=0.8, every node/.style={minimum size=0.7cm}]
      \node (A) at (0, 0) {$c_1$};
      \node (B) at (3, 0) {$c_2$};
      \node (C) at (1.5, 2.6) {$c_3$};
      \draw[draw] (A) circle [radius=0.45];
      \draw[draw] (B) circle [radius=0.45];
      \draw[draw] (C) circle [radius=0.45];
    
      \draw[-] (A) to node[midway, sloped, fill=white, inner sep=1pt] {$e_{12}$} (B);
      \draw[-] (B) to node[midway, sloped, fill=white, inner sep=1pt] {$e_{11}$} (C);
      \draw[-] (A) to node[midway, sloped, fill=white, inner sep=1pt] {$e_{13}$} (C);
    
      \draw[-] (A) [loop left] to node[left] {$e_{11}$} (A);
      \draw[-] (B) [loop right] to node[right] {$e_{22}$} (B);
      \draw[-] (C) [loop above] to node[above] {$e_{33}$} (C);
    \end{tikzpicture}
    \caption{Schematic representation of relevant variables tracked in the 3-state $q$-voter model with annealed disorder.} 
    \label{fig:annealed:scheme}
\end{figure}
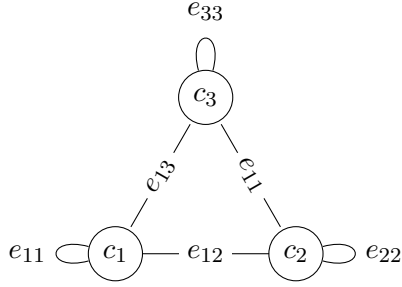
In the case of annealed disorder, there are $S$ distinct voter classes, as the class of a voter is equivalent to its state. We will approximate the probabilities $P[s'|s]$ in Eq. \eqref{eqn:general:pa:annealed} using the concentration of edge classes:
\begin{eqnarray}
    P[s'|s] = \frac{e_{ss'}}{\sum_{\sigma}e_{s\sigma}} := \theta_s^{s'},
    \label{eqn:theta:annealed}
\end{eqnarray}
where $e_{ss'} := E_{ss'}/E$; $E$ is the total number of directed edges in the graph (twice the number of edges in the underlying undirected graph) and $E_{ss'}$ is the number of directed edges starting at a voter in state $s$ and ending at a voter in state $s'$. Of course $E_{ss'}$ and therefore $e_{ss'}$, can change over time, so we should actually write $E_{ss'}(t)$ and $e_{ss'}(t)$, but for the sake of brevity we omit the time dependence from the notation. Analogously to state concentrations, we assume that edge concentrations $e_{ss'}(t)$ convergence to their expected values.
There are in total $S(S-1)/2 + S$ distinct edge classes, e.g., 6 classes for $s=3$, as shown in Fig.~\ref{fig:annealed:scheme}. Since the underlying graphs are undirected, $e_{ss'} = e_{s's}$ always holds. Furthermore, only $S(S-1)/2 + S - 1$ edge concentrations are independent, as their sum must equate to one: $\sum_{s}\sum_{s'}e_{ss'} = 1$.

Introducing edge concentrations was helpful to approximate the probability of selecting neighbors in specific states, but we now need to track the time evolution of $e_{ss'}$. We can write the general equation for the time derivatives of the concentration of edges $e_{ss'}$:
\begin{eqnarray}
    \frac{d e_{ss'}}{dt} = \frac{1}{\langle k \rangle}\sum_{\sigma \neq \sigma'} c_{\sigma} \sum_{\boldsymbol{k}}P(\boldsymbol{k}| \sigma) f^{\sigma\rightarrow \sigma'}(\boldsymbol{k}) \Delta E_{ss'}\big|^{\sigma\rightarrow \sigma'}(\boldsymbol{k}),
    \label{eqn:de_dt_pa:annealed}
\end{eqnarray}
where $\langle k \rangle$ is the global average node degree, $\Delta E_{ss'}\big|^{\sigma\rightarrow \sigma'}$ is the change in $E_{ss'}$ occurring when a voter flips its state from $\sigma$ to $\sigma'$, given its neighborhood vector $\boldsymbol{k}$. The components of $\boldsymbol{k}$ are the numbers of neighbors in each state, i.e. $\boldsymbol{k}= [k_1, ..., k_S]$ and $k = \sum_sk_s$ is the node degree (total number of neighbors). When a voter flips from $\sigma$ to $\sigma'$, all of its outgoing edges change from $(\sigma, \cdot)$ to $(\sigma', \cdot)$ and all of its incoming edges change from $(\cdot, \sigma)$ to $(\cdot, \sigma')$, where $\cdot$ denotes an arbitrary state. The number of neighbors in each state directly corresponds to the number of outgoing edges ending at this state and the number of incoming edges originating at this state (because every edge has an oppositely directed counterpart). Hence the change of $E_{ss'}$ during a single voter update is given by
\begin{eqnarray}
    \Delta E_{ss'}\big|^{\sigma \rightarrow \sigma'}(\boldsymbol{k}) = k_{s'}(\delta_{\sigma' s} - \delta_{\sigma s}) + k_s(\delta_{\sigma' s'}-\delta_{\sigma s'}).
\label{eqn:delta:e:annealed}
\end{eqnarray}
For example, when a voter in state $\sigma=1$ changes its state to any other state $\sigma'$, the number of edges $E_{11}
$ between voters in state $1$ changes by $\Delta E_{11}\big|^{1 \rightarrow \sigma'}(\boldsymbol{k}) = -2k_1$. The flipping probability expressed in terms of the neighborhood vector, $f^{\sigma \rightarrow \sigma'}(\boldsymbol{k})$, takes the form:
\begin{eqnarray}
    f^{\sigma \rightarrow \sigma'}(\boldsymbol{k}) = (1-p)\frac{k_{\sigma'}!(k-q)!}{k!(k_{\sigma'} - q)!}\boldsymbol{1}_{k_{\sigma'} \geq q} + \frac{p}{S-1}\frac{k_\sigma!(k-q)!}{k!(k_\sigma - q)!}\boldsymbol{1}_{k_\sigma \geq q}.
    \label{eqn:flipping:annealed}
\end{eqnarray}
Notice that the flipping probability used in the calculation of edge concentrations is different from the one used in Eq. \eqref{eqn:general:pa:annealed}. This stems from the fact that tracking the change in the number of edges requires information about the states of all neighbors. The flipping probability thus needs to be conditioned on this information, since the event of constructing a $q$-panel is dependent on the neighborhood vector. To arrive at a usable form of Eq.~\eqref{eqn:de_dt_pa:annealed} we need to calculate the innermost sum of this equation. Since we assumed that the states of the neighbors are i.i.d. the probability that a voter in state $\sigma$ has $k_{\sigma'}$ neighbors in state $\sigma'$ is of the binomial form:
\begin{eqnarray}
    P(k_{\sigma'}|\sigma, k) = \binom{k}{k_{\sigma'}}(\theta_\sigma^{\sigma'})^{k_{\sigma'}}(1-\theta_\sigma^{\sigma'})^{k - k_{\sigma'}}.
    \label{eqn:binomial:annealed}
\end{eqnarray}
Plugging Eqs. \eqref{eqn:flipping:annealed} and \eqref{eqn:delta:e:annealed} into the innermost sum of Eq.~\eqref{eqn:de_dt_pa:annealed}, we arrive at: 
\begin{eqnarray}
    \sum_{\boldsymbol{k}}P(\boldsymbol{k}| \sigma) f^{\sigma\rightarrow \sigma'}(\boldsymbol{k}) \Delta E_{ss'}|^{\sigma\rightarrow \sigma'}(\boldsymbol{k}) &=& \sum_{\boldsymbol{k}}P(\boldsymbol{k}| \sigma) \nonumber \\ 
    &\times& \underbrace{\Bigl((1-p)\frac{k_{\sigma'}!(k-q)!}{k!(k_{\sigma'}-q)!}\boldsymbol{1}_{k_{\sigma'}\geq q} + \frac{p}{S-1}\frac{k_{\sigma}!(k-q)!}{k!(k_{\sigma}-q)!}\boldsymbol{1}_{k_{\sigma}\geq q}\Bigr)}_{\text{R1}} \nonumber \\ 
    &\times& \underbrace{\Bigl(k_{s'}(\delta_{\sigma' s} - \delta_{\sigma s}) + k_s(\delta_{\sigma' s'}-\delta_{\sigma s'})\Bigr)}_{\text{R2}}.
\label{eqn:part1:annealed}
\end{eqnarray}
After performing multiplication of terms R1 and R2 from Eq.~\eqref{eqn:part1:annealed}, we end up with terms that belong to one of the two types (omitting $(1-p)$ and $p/(S-1)$ factors), the first one for $s=s'$:
\begin{eqnarray}
     && \sum_{\boldsymbol{k}}P(\boldsymbol{k}| \sigma)\frac{k_s!(k-q)!}{k!(k_s-q)!}\boldsymbol{1}_{k_s\geq q}k_s =\sum_{k\in\mathbb{N}}P(k| \sigma)\sum_{k_s=q}^k P(k_s | \sigma, k)\frac{k_s!(k-q)!}{k!(k_s-q)!}k_s\nonumber \\
     && =\sum_{k\in\mathbb{N}}P(k| \sigma)\sum_{k_s=q}^k\binom{k}{k_s}\left(\theta_\sigma^s\right)^{k_s} \left(1-\theta_\sigma^s\right)^{k-k_s}\frac{k_s!(k-q)!}{k!(k_s-q)!}k_s \nonumber \\
     && =\sum_{k\in\mathbb{N}}P(k| \sigma) \left(\theta_\sigma^s\right)^q\left[\left( k-q \right)\theta_\sigma^s + q \right] = \left(\theta_\sigma^s\right)^q\left[\left( \langle k \rangle_{\sigma} -q \right)\theta_\sigma^s + q \right]
     \label{eqn:uniform:sum:annealed}
 \end{eqnarray}
and the other for $s \ne s'$:
\begin{eqnarray}
    && \sum_{\boldsymbol{k}}P(\boldsymbol{k}| \sigma)\frac{k_{s}!(k-q)!}{k!(k_{s}-q)!}\boldsymbol{1}_{k_{s}\geq q}k_{s'} = \sum_{k\in\mathbb{N}}P(k| \sigma)  \sum_{k_{s}=q}^{k}P(k_{s}| \sigma, k) \frac{k_{s}!(k-q)!}{k!(k_{s}-q)!} \sum_{k_{s'}=0}^{k-k_{s}}P(k_{s'}| \sigma, k, k_{s}) k_{s'} \nonumber \\
     && =\sum_{k\in\mathbb{N}}P(k| \sigma)\sum_{k_{s}=q}^k\binom{k}{k_{s}}\left(\theta_{\sigma}^s\right)^k_{s} \left(1-\theta_{\sigma}^s\right)^{k-k_{s}}\frac{k_{s}!(k-q)!}{k!(k_{s}-q)!} \nonumber \\ 
     && \times\underbrace{\sum_{k_{s'}=0}^{k-k_{s}}\binom{k-k_{s}}{k_{s'}} \left(\frac{\theta_{\sigma}^{s'}}{1-\theta_{\sigma}^s}\right)^{k_s'} \left(1-\frac{\theta_{\sigma}^{s'}}{1-\theta_{\sigma}^s}\right)^{k-k_{s}-k_{s'}}k_{s'}}_{ = \left(k - k_{s}\right)\theta_{\sigma}^{s'}/\left(1-\theta_{\sigma}^s\right)} \label{eqn:mixed:sum:annealed}\\
     && = \sum_{k\in\mathbb{N}}P(k| \sigma)\frac{\theta_{\sigma}^{s'}}{1-\theta_{\sigma}^s}\left(\theta_{\sigma}^s\right)^q k - \sum_{k\in\mathbb{N}}P(k| \sigma)\frac{\theta_{\sigma}^{s'}}{1-\theta_{\sigma}^s}\left(\theta_{\sigma}^s\right)^q\left[\left( k-q \right)\theta_{\sigma}^s + q \right] = \theta_{\sigma}^{s'} \left(\theta_{\sigma}^s\right)^q \left(\langle k \rangle_{\sigma} - q\right).\nonumber 
\end{eqnarray}

Now, plugging Eqs \eqref{eqn:uniform:sum:annealed} and \eqref{eqn:mixed:sum:annealed} to Eq. \eqref{eqn:de_dt_pa:annealed}, we obtain the final evolution equations for edges in annealed systems:
\begin{eqnarray}
    \frac{d e_{ss'}}{dt} &=& \frac{1}{\langle k \rangle}\sum_{\substack{\sigma \neq \sigma'}} c_{\sigma}\sum_{\boldsymbol{k}}P(\boldsymbol{k}| \sigma) f^{\sigma \rightarrow \sigma'}(\boldsymbol{k}) \Delta E_{ss'}^{\sigma\rightarrow \sigma'}(\boldsymbol{k}) = \frac{1}{\langle k \rangle}\sum_{\substack{\sigma \neq \sigma'}} c_{\sigma} \Bigl[(1-p) \left(\theta_{\sigma}^{\sigma'}\right)^q \nonumber \\ 
    &\times& \Bigl\{\left(\left[\langle k \rangle_{\sigma} - q \right]\theta_{\sigma}^{s'} + q\delta_{\sigma' s'} \right)\left(\delta_{\sigma' s} - \delta_{\sigma s} \right) +\left(\left[\langle k \rangle_{\sigma} - q \right]\theta_\sigma^s + q\delta_{\sigma' s}\right)\left(\delta_{\sigma' s'} - \delta_{\sigma s'} \right) \Bigr\} \nonumber \\
    &+&\frac{p}{S-1}\left(\theta_\sigma^\sigma\right)^q \Bigl\{\left(\left[\langle k \rangle_{\sigma} - q \right]\theta_\sigma^{s'} + q\delta_{\sigma s'} \right)\left(\delta_{\sigma' s} - \delta_{\sigma s} \right) \nonumber \\ 
    &+& \left(\left[\langle k \rangle_{\sigma} - q \right]\theta_\sigma^s + q\delta_{\sigma s}\right)\left(\delta_{\sigma' s'} - \delta_{\sigma s'} \right) \Bigr\} \Bigr].
   \label{eqn:final:annealed}
\end{eqnarray}
Where the average node degree $\langle k \rangle_\sigma$ of voters in state $\sigma$ can be inferred from the state and edge concentrations:
\begin{eqnarray}
    \langle k \rangle_\sigma = \frac{\sum_{\sigma'}e_{\sigma\sigma'}}{c_\sigma}\langle k \rangle.
    \label{eqn:ave:degree:annealed}
\end{eqnarray}
The formulas for time derivatives of state concentrations $c_s$ are obtained by plugging Eqs~\eqref{eqn:general:pa:annealed} and \eqref{eqn:theta:annealed} into Eq.~\eqref{eqn:general}:
\begin{eqnarray}
    \frac{dc_s}{dt} = \sum_{\sigma\neq\sigma'}(\delta_{s\sigma'} - \delta_{s\sigma})c_{\sigma} \left[(1-p)\left(\theta_\sigma^{\sigma'}\right)^q + \frac{p}{S-1}\left(\theta_\sigma^\sigma\right)^q\right].
    \label{eqn:final:annealed:c}
\end{eqnarray}
\subsection{Quenched disorder}

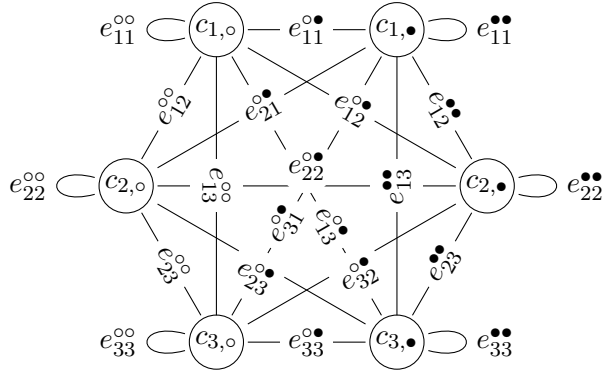
\begin{figure}[htbp]
    \begin{tikzpicture}[scale=0.8, every node/.style={minimum size=0.7cm}]
      \node (c1c) at (120:3) {$c_{1,\circ}$};
      \node (c2c) at (180:3) {$c_{2,\circ}$};
      \node (c3c) at (240:3) {$c_{3,\circ}$};
      \node (c1a) at (60:3) {$c_{1,\bullet}$};
      \node (c2a) at (0:3) {$c_{2,\bullet}$};
      \node (c3a) at (300:3) {$c_{3,\bullet}$};
      
      \draw[draw] (c1a) circle [radius=0.45];
      \draw[draw] (c2a) circle [radius=0.45];
      \draw[draw] (c3a) circle [radius=0.45];
      \draw[draw] (c1c) circle [radius=0.45];
      \draw[draw] (c2c) circle [radius=0.45];
      \draw[draw] (c3c) circle [radius=0.45];
    
      \draw[-] (c1c) [loop left] to node[left] {$e_{11}^{\circ\circ}$} (c1c);
      \draw[-] (c1c) to node[midway, sloped, fill=white, inner sep=1pt] {$e_{12}^{\circ\circ}$} (c2c);
      
      \draw[-] (c1c) to node[midway, sloped, fill=white, inner sep=1pt] {$e_{11}^{\circ\bullet}$} (c1a);
      
      \draw[-] (c1c) to node[circle, pos=0.65, sloped, fill=white, inner sep=1pt] {$e_{13}^{\circ\bullet}$} (c3a);

      \draw[-] (c2c) [loop left] to node[left] {$e_{22}^{\circ\circ}$} (c2c);
      \draw[-] (c2c) to node[midway, sloped, fill=white, inner sep=1pt] {$e_{23}^{\circ\circ}$} (c3c);
      \draw[-] (c2c) to node[midway, sloped, fill=white, inner sep=1pt] {$e_{21}^{\circ\bullet}$} (c1a);

      \draw[-] (c3c) [loop left] to node[left] {$e_{33}^{\circ\circ}$} (c3c);
      \draw[-] (c3c) to node[circle, pos=0.35, sloped, fill=white, inner sep=1pt] {$e_{31}^{\circ\bullet}$} (c1a);
      
      \draw[-] (c3c) to node[midway, sloped, fill=white, inner sep=1pt] {$e_{33}^{\circ\bullet}$} (c3a);

      \draw[-] (c1a) [loop right] to node[right] {$e_{11}^{\bullet\bullet}$} (c1a);
      \draw[-] (c1a) to node[midway, sloped, fill=white, inner sep=1pt] {$e_{12}^{\bullet\bullet}$} (c2a);

      \draw[-] (c2a) [loop right] to node[right] {$e_{22}^{\bullet\bullet}$} (c2a);
      \draw[-] (c2a) to node[midway, sloped, fill=white, inner sep=1pt] {$e_{23}^{\bullet\bullet}$} (c3a);
      
      \draw[-] (c3a) [loop right] to node[right] {$e_{33}^{\bullet\bullet}$} (c3a);
     
      \draw[-] (c2c) to node[circle, midway, sloped, fill=white, above=-3pt, inner sep=1pt] {$e_{22}^{\circ\bullet}$} (c2a);
      \draw[-] (c1c) to node[midway, sloped, fill=white, inner sep=1pt] {$e_{12}^{\circ\bullet}$} (c2a);
      \draw[-] (c2c) to node[midway, below=-5pt, sloped, fill=white, inner sep=1pt] {$e_{23}^{\circ\bullet}$} (c3a);
      \draw[-] (c3c) to node[midway, below=-5pt, sloped, fill=white, inner sep=1pt] {$e_{32}^{\circ\bullet}$} (c2a);
      \draw[-] (c1a) to node[midway, sloped, fill=white, inner sep=1pt] {$e_{13}^{\bullet\bullet}$} (c3a);
      \draw[-] (c1c) to node[midway, sloped, fill=white, inner sep=1pt] {$e_{13}^{\circ\circ}$} (c3c);
      
    \end{tikzpicture}
    \caption{Schematic representation of relevant variables tracked in the 3-state $q$-voter model with conformists $\circ$ and anticonformists $\bullet$ in quenched disorder.}
    \label{fig:quenched:scheme}
\end{figure}
If the $q$-voter model is implemented with the quenched disorder, each voter $v$ is ascribed with a quenched variable $\eta_v$. Thus, given $S$ possible states, there are in total $2S$ voter classes, as each of the voter can be either a conformist $\circ$ or an anticonformist $\bullet$. This leads to $2S^2 + S$ different edge classes, e.g. $21$ edge variables for $s=3$, as shown in Fig.~\ref{fig:quenched:scheme}. The concentrations of voters in each class are described by the variables $e_{ss'}^{\eta\eta'} := E_{ss'}^{\eta\eta'}/E$, where $E_{ss'}^{\eta\eta'} = E_{s's}^{\eta'\eta}$ is the number of directed edges from a voter in state $s$ of type $\eta$ to a voter in state $s'$ and of type $\eta'$. The probability $P\left[(s', \eta') | (s, \eta)\right]$ that a randomly selected neighbor of a voter $(s, \eta)$ is $(s', \eta')$ can be expressed as:
\begin{eqnarray}
    P\left[(s', \eta') | (s, \eta) \right] = \frac{e_{ss'}^{\eta\eta'}}{\sum_{\sigma} e_{s\sigma}^{\eta\circ} + e_{s\sigma}^{\eta\bullet}} := \theta_{s\eta}^{s'\eta'},
\end{eqnarray}
and if only the state of the neighbor is of interest, we can write:
\begin{eqnarray}
    P\left[s' | (s, \eta) \right] = \theta_{s\eta}^{s'\circ} + \theta_{s\eta}^{s'\bullet} := \theta_{s\eta}^{s'}.
    \label{eqn:theta:quenched}
\end{eqnarray}
Since the fraction (or its expected value) of conformists and anticonformists in the system is fixed and given by
\begin{eqnarray}
    c_\circ &:=& \sum_{s} c_{s,\circ} = (1-p), \nonumber \\
    c_\bullet &:=& \sum_{s} c_{s,\bullet} = p,
\end{eqnarray}
the number of independent edge concentrations can be reduced using the following conditions: 
\begin{eqnarray}
    e^{\circ\circ} &:=& \sum_{s}\sum_{s'}  e_{ss'}^{\circ\circ} = (1-p)^2, \\
    e^{\circ\bullet} &:=& \sum_{s}\sum_{s'}  e_{ss'}^{\circ\bullet} = (1-p)p,\\
    e^{\bullet\bullet} &:=& \sum_{s}\sum_{s'}  e_{ss'}^{\bullet\bullet} = p^2,
\end{eqnarray}
leading to $2S^2 + S - 3$ independent edge concentrations. We can formulate the general equation for the evolution of edge concentrations as
\begin{eqnarray}
    \frac{d e_{ss'}^{\eta\eta'}}{dt} &=& \frac{1}{\langle k \rangle}\sum_{\beta\in\{\circ, \bullet\}}\sum_{\sigma \neq \sigma'} c_{\sigma,\beta} \sum_{\boldsymbol{k}}P(\boldsymbol{k}| (\sigma,\beta)) f_\beta^{\sigma\rightarrow \sigma'}(\boldsymbol{k}) \Delta E_{ss'}^{\eta\eta'}\big|_\beta^{\sigma \rightarrow \sigma'}(\boldsymbol{k}).
    \label{eqn:dedt:pa:general:quenched}
\end{eqnarray}

In the presence of quenched disorder, the flipping probabilities are now dependent on the state $s$ as well as the type $\eta$ of the target voter:
\begin{eqnarray}
    f_\eta^{s\rightarrow s'}(\boldsymbol{k}) &=& \delta_{\eta\circ}\frac{k_{s'}!(k-q)!}{k!(k_{s'}-q)!}\boldsymbol{1}_{k_{s'}\geq q} + \frac{\delta_{\eta\bullet}}{S-1}\frac{k_s!(k-q)!}{k!(k_s-q)!}\boldsymbol{1}_{k_s\geq q},
    \label{eqn:flipping:quenched}
\end{eqnarray}
with the neighborhood vector $\boldsymbol{k} = \left[k_1^\bullet, ..., k_S^\bullet, k_1^\circ, ..., k_S^\circ \right]$, where $k_s^\bullet$ and $k_s^\circ$  are respectively the number of anticonformist and conformist neighbors in state $s$, and $k_s = k_s^\bullet + k_s^\circ$ is the total number of neighbors in state $s$. 

The elementary change in the number of edges $E_{ss'}^{\eta\eta'}$ occurring when a voter of type $\beta$ and neighborhood vector $\boldsymbol{k}$ flips its state from $\sigma$ to $\sigma'$:
\begin{eqnarray}
    \Delta E_{ss'}^{\eta\eta'}\big|_\beta^{\sigma\rightarrow \sigma'}(\boldsymbol{k}) &=& \delta_{\beta\eta}k_{s'}^{\eta'}(\delta_{\sigma's} - \delta_{\sigma s}) + \delta_{\beta\eta'}k_s^{\eta}(\delta_{\sigma's'}-\delta_{\sigma s'}).
\label{eqn:delta:e:quenched}
\end{eqnarray}
The probability that a voter in class $(\sigma, \beta)$ with $k$ total neighbors has $k_s^\eta$ neighbors in state $s$ and of type $\eta$ is
\begin{eqnarray}
    P(k_s^\eta|(\sigma, \beta), k) = \binom{k}{k_s^\eta}(\theta_{\sigma\beta}^{s\eta})^{k_s^\eta}(1-\theta_{\sigma \beta}^{s\eta})^{k - k_s^\eta}.
    \label{eqn:binomial:quenched}
\end{eqnarray}
Now, using Eqs \eqref{eqn:flipping:quenched}, \eqref{eqn:delta:e:quenched} and \eqref{eqn:binomial:quenched} we can express the innermost sum of Eq.~\eqref{eqn:dedt:pa:general:quenched} as
\begin{eqnarray}
   && \sum_{\boldsymbol{k}}P(\boldsymbol{k}| (\sigma, \beta)) f_\beta^{\sigma\rightarrow \sigma'}(\boldsymbol{k}) \Delta E_{ss'}^{\eta\eta'}\big|_\beta^{\sigma \rightarrow \sigma'}(\boldsymbol{k}) = \sum_{\boldsymbol{k}}P(\boldsymbol{k}| (\sigma, \beta)) \nonumber \\
   && \times \underbrace{\left(\delta_{\beta\circ}\frac{k_{\sigma'}!(k-q)!}{k!(k_{\sigma'}-q)!}\boldsymbol{1}_{k_{\sigma'}\geq q} + \frac{\delta_{\beta\bullet}}{S-1}\frac{k_\sigma!(k-q)!}{k!(k_\sigma-q)!}\boldsymbol{1}_{k_\sigma\geq q}\right)}_{\text{R1}} \nonumber \\ && \times \underbrace{\Bigl(\delta_{\beta\eta}k_{s'}^{\eta'}(\delta_{\sigma's} - \delta_{\sigma s}) + \delta_{\beta\eta'}k_s^\eta(\delta_{\sigma's'}-\delta_{\sigma s'})\Bigr)}_{\text{R2}}.
\label{eqn:part1:quenched}
\end{eqnarray}
To obtain usable equations, we need to simplify the terms that arise after multiplying R1 with R2 in Eq.~\eqref{eqn:part1:quenched}, they can be either equivalent to (for $s=s'$):
\begin{eqnarray}
   && \sum_{\boldsymbol{k}} P(\boldsymbol{k}| (\sigma, \beta)) \frac{k_s!(k-q)!}{ k!(k_s-q)!}\boldsymbol{1}_{k_s\geq q} k^\eta_s {=} \sum_{k \in \mathbb{N}}P(k|(\sigma, \beta){)} \times\sum_{k_s=q}^k\binom{k}{k_s}\left(\theta_\sigma^s\right)^k_s \left(1-\theta_\sigma^s\right)^{k-k_s}\frac{k_s!(k-q)!}{ k!(k_s-q)!} \nonumber \\
    && \times \underbrace{\sum_{k^\eta_s=0}^{k_s}\binom{k_s}{k^\eta_s}\left(\frac{\theta_\sigma^{s\eta}}{\theta_\sigma^s}\right)^{k_s^\eta} \left(1-\frac{\theta_\sigma^{s\eta}}{\theta_\sigma^s}\right)^{k_s-k_s^\eta}k_s^\eta}_{ = k_s\theta_\sigma^{s\eta}/\theta_\sigma^s} = \sum_{k \in \mathbb{N}}P(k|(\sigma, \beta){)}\frac{\theta_\sigma^{s\eta}}{\theta_\sigma^s}\left(\theta_\sigma^s\right)^q\left[\left( k-q \right)\theta_\sigma^s + q \right] \nonumber \\
    && = \theta_\sigma^{s\eta}\left(\theta_\sigma^s\right)^{q-1}\left[\left( \langle k \rangle_{\sigma, \beta} -q \right)\theta_\sigma^s + q \right]
\end{eqnarray}
or (for $s \ne s'$): 
\begin{eqnarray}
    &&\sum_{\boldsymbol{k}}P(\boldsymbol{k}| (\sigma, \beta)) \frac{k_s!(k-q)!}{ k!(k_s-q)!}\boldsymbol{1}_{k_s\geq q} k^\eta_{s'} = \sum_{k \in \mathbb{N}}P(k|(\sigma, \beta){)} \sum_{k_s=q}^k\binom{k}{k_s}\left(\theta_\beta^s\right)^k_s \left(1-\theta_\beta^s\right)^{k-k_s}\frac{k_s!(k-q)!}{ k!(k_s-q)!} \nonumber \\
    && \underbrace{\sum_{k^\eta_{s'}=0}^{k-k_s}\binom{k-k_s}{k^\eta_{s'}}\left(\frac{\theta_\beta^{s'\eta}}{1-\theta_\beta^s}\right)^{k_s^\eta} \left(1-\frac{\theta_\beta^{s'\eta}}{1-\theta_\beta^s}\right)^{k-k_s-k_{s'}^\eta}k_{s'}^\eta}_{=(k-k_s)\theta_\beta^{s'\eta}/(1-\theta_\beta^s)} \nonumber \\ 
    && = \sum_{k \in \mathbb{N}}P(k|(\sigma, \beta){)}\frac{\theta_\beta^{s'\eta}}{1-\theta_\beta^s}\left(\theta_\beta^s\right)^q (k - \left[\left(k - q \right)\theta_\beta^s + q \right]) = \theta_\beta^{s'\eta}\left(\theta_\beta^s\right)^q \left( \langle k \rangle_{\sigma, \beta} - q \right).
\end{eqnarray}

This finally yields a set of equations describing the time evolution of edge-class concentrations, which enables a numerical analysis of the quenched system within the pair approximation:

\begin{eqnarray}
    \frac{d e_{ss'}^{\eta\eta'}}{dt} &=& \frac{1}{\langle k \rangle}\sum_{\beta\in\{\circ, \bullet\}}\sum_{\sigma \neq \sigma'} c_{\sigma, \beta} \Bigl[ \delta_{\beta \circ} \left(\theta_{\sigma\beta}^{\sigma'} \right)^q \Bigl\{ \delta_{\beta \eta} \left(\left[\langle k \rangle_{\sigma \beta} - q \right] + \tfrac{q}{\theta_{\sigma \beta}^{s'}}\delta_{\sigma' s'} \right)\theta_{\sigma \beta}^{{s'}\eta'}\left(\delta_{\sigma' s} - \delta_{\sigma s} \right) \nonumber \\ 
    &+& \delta_{\beta \eta'} \left(\left[\langle k \rangle_{\sigma \beta} - q \right] + \tfrac{q}{\theta_{\sigma \beta}^{s}}\delta_{\sigma' s} \right)\theta_{\sigma \beta}^{s\eta}\left(\delta_{\sigma' s'} - \delta_{\sigma s'} \right)\Bigr\} \nonumber \\
    &+& \frac{\delta_{\beta \bullet}}{S-1} \left(\theta_{\sigma \beta}^\sigma\right)^q \Bigl\{ \delta_{\beta \eta}\left(\left[\langle k \rangle_{\sigma \beta} - q \right] + \tfrac{q}{\theta_{\sigma \beta}^{s'}}\delta_{\sigma s'} \right)\theta_{\sigma \beta}^{{s'}\eta'}\left(\delta_{\sigma' s} - \delta_{\sigma s} \right) \nonumber \\ 
    &+& \delta_{\beta\eta'} \left(\left[\langle k \rangle_{\sigma\beta} - q \right] + \tfrac{q}{\theta_{\sigma \beta}^{s}}\delta_{\sigma s} \right)\theta_{\sigma \beta}^{s\eta}\left(\delta_{\sigma' s'} - \delta_{\sigma s'} \right) \Bigr\}\Bigr].
    \label{eqn:final:quenched}
\end{eqnarray}

In the quenched scenario, the average node degree $\langle k \rangle_{\sigma \beta}$ of voters in state $\sigma$ and type $\beta$ given by:
\begin{eqnarray}
    \langle k \rangle_{\sigma \beta} = \frac{\sum_{s}e_{\sigma s}^{\beta\circ}+e_{\sigma s}^{\beta\bullet}}{c_{\sigma,\beta}}\langle k \rangle.
    \label{eqn:ave:degree:quenched}
\end{eqnarray}

For completeness, we also provide the explicit formula for the evolution of voter class concentrations $c_{s, \eta}$, obtained from specifying the general form Eq.~\eqref{eqn:general} with Eqs~\eqref{eqn:general:pa:quenched} and \eqref{eqn:theta:quenched}:
\begin{eqnarray}
    \frac{dc_{s,\eta}}{dt} &=& \sum_{\sigma\neq\sigma'}(\delta_{s\sigma'} - \delta_{s\sigma})c_{\sigma, \eta} \Bigl[\delta_{\eta\circ}\left(\theta_{\sigma\eta}^{\sigma'}\right)^q + \frac{\delta_{\eta\bullet}}{S-1}\left(\theta_{\sigma\eta}^\sigma\right)^q\Bigr].
    \label{eqn:final:quenched:c}
\end{eqnarray}

\subsection{State-degree correlation}

Having calculation-ready formulas for the time derivatives of edge class concentrations as well as voter class concentrations, Eqs. \eqref{eqn:final:annealed} and \eqref{eqn:final:annealed:c} for the annealed approach, Eqs. \eqref{eqn:final:quenched} and \eqref{eqn:final:quenched:c} for the quenched approach, we can describe the evolution of the system within the pair approximation. In both the annealed and quenched disorder, the evolution does not depend on the degree distribution, but only on the average node degree within each voter class, described with Eqs. \eqref{eqn:ave:degree:annealed} and \eqref{eqn:ave:degree:quenched}. It should be noted that the result showing that the degree distribution does not influence the dynamics has previously been obtained for binary $q$-voter models with annealed independence \cite{Jedrzejewski2017PairNetworks}, as well as with quenched independence and quenched anticonformity \cite{Jedrzejewski2020NonlinearPerspective}. However, such a simplification, where the outcome depends only on the average degree, arises only in the model where neighbors in the $q$-panel are selected without repetitions. If repetitions are allowed, the dynamical equations within PA depend on the full degree distribution \cite{Peralta2018, Vieira2020PairModel, Ramirez2024OrderingModels}.

We will now show that if initially there are no correlations between the average node degree and the class of the voter, i.e., $\langle k \rangle_{s, \eta} = \langle k \rangle$ for every $s$ and $\eta$, such correlations will not arise in the system, generalizing the result for the binary version of the model~\cite{Jedrzejewski2022PairNetworks}. Here, we present calculations for the quenched scenario, but analogous considerations can be conducted for the simpler annealed case. Let us start by considering how the numerator of Eq. \eqref{eqn:ave:degree:quenched} changes w.r.t time:

\begin{eqnarray}
    \frac{d}{dt}&\Bigl(&\sum_{s'}\sum_{\eta'} e_{ss'}^{\eta \eta'} \Bigr) = \frac{1}{\langle k \rangle} \sum_{s'}\sum_{\eta'} \sum_\beta\sum_{\sigma \neq \sigma'} c_{\sigma,\beta} \delta_{\beta \circ} \left(\theta_{\sigma \beta}^{\sigma'}\right)^q \nonumber \\ 
    &\times& \Bigl\{ \delta_{\beta\eta} \left[\left(\langle k \rangle_{\sigma\beta} - q \right) + \tfrac{q}{\theta_{\sigma \beta}^{s'}}\delta_{\sigma' s'} \right]\theta_{\sigma\beta}^{s'\eta'}\left(\delta_{\sigma' s} - \delta_{\sigma s} \right) \nonumber \\ 
    &+& \delta_{\beta \eta'} \left[\left(\langle k \rangle_{\sigma \beta} - q \right) + \tfrac{q}{\theta_{\sigma \beta}^{s}}\delta_{\sigma' s} \right]\theta_{\sigma\beta}^{s\eta}\left(\delta_{\sigma' s'} - \delta_{\sigma s'} \right)\Bigr\} + c_{\sigma,\beta}(\delta_{\beta \bullet}...) \nonumber \\
     &=& \frac{1}{\langle k \rangle} \sum_\beta\sum_{\sigma \neq \sigma'} c_{\sigma,\beta} \delta_{\beta \circ}\left(\theta_{\sigma \beta}^{\sigma'}\right)^q \Bigl\{\delta_{\beta \eta}\sum_{s'}\sum_{\eta'}\left[ (\langle k \rangle_{\sigma\beta} - q) + \tfrac{q}{\theta_{\sigma\beta}^{s'}} \delta_{\sigma's'}\right]\theta_{\sigma\beta}^{s' \eta'}(\delta_{\sigma's} - \delta_{\sigma s})\nonumber \\
     &+&\left[ (\langle k \rangle_{\sigma,\beta} - q) + \tfrac{q}{\theta_{\sigma\beta}^{s}} \delta_{\sigma's} \right] \theta_{\sigma\beta}^{s\eta}\underbrace{\sum_{s'}(\delta_{\sigma' s'} - \delta_{\sigma s'})}_{=0} \sum_{\eta'}\delta_{\beta \eta'}\Bigr\} + \sum_{s'}\sum_{\eta'}c_{\sigma,\beta}(\delta_{\beta \bullet}...) \nonumber \\
     &=& \frac{1}{\langle k \rangle} \sum_{\beta}\sum_{\sigma \neq \sigma'} c_{\sigma,\beta} \delta_{\beta \circ}\left(\theta_{\sigma\beta}^{\sigma'}\right)^q\delta_{\beta\eta} (\delta_{\sigma's} - \delta_{\sigma s}) \nonumber \\ 
     &\times& \Bigl[(\langle k \rangle_{\sigma\beta} - q)\underbrace{\sum_{s'}\sum_{\eta'}\theta_{\sigma \beta}^{s'\eta'}}_{=1} + q \underbrace{\sum_{s'}\delta_{\sigma's'}\sum_{\eta'}\tfrac{\theta_{\sigma\beta}^{s'\eta'}}{\theta_{\sigma\beta}^{s'}}}_{=1}\Bigr] + \sum_{s'}\sum_{\eta'}c_{\sigma,\beta}(\delta_{\beta \bullet}...) \nonumber \\
      &=& \frac{1}{\langle k \rangle} \sum_{\sigma \neq \sigma'} c_{\sigma,\eta} \delta_{\eta \circ}\left(\theta_{\sigma \eta}^{\sigma'}\right)^q (\delta_{\sigma' s} - \delta_{\sigma s})\langle k \rangle_{\sigma \eta} + c_{\sigma,\eta}(\delta_{\eta \bullet}...) \nonumber \\ &=& \frac{1}{\langle k \rangle} \sum_{\sigma \neq \sigma'} (\delta_{\sigma's} - \delta_{\sigma s})c_{\sigma,\eta}\left[\delta_{\eta \circ}\left(\theta_{\sigma \eta}^{\sigma'}\right)^q + \frac{\delta_{\eta \bullet}}{S-1} \left(\theta_{\sigma \eta}^\sigma\right)^q\right] \langle k \rangle_{\sigma \eta}
    \label{eqn:sum_de}
\end{eqnarray}
The term $(\delta_{\beta\bullet}...)$ stands for the terms corresponding to anticonformists, but since the calculations are analogous, we omit it for clarity. Now note that if $\forall_{s, \eta} \langle k \rangle_{s, \eta} = \langle k \rangle$, then $\frac{d}{dt}\left( \sum_{s'}\sum_{\eta'} e_{s s'}^{\eta \eta'} \right)$ is equal to $\frac{dc_{s,\eta}}{dt}$, using the quotient rule it is straightforward to show that $\frac{d}{dt}\frac{\sum_{s'}\sum_{\eta'} e_{s s'}^{\eta \eta'}}{c_{s \eta}} = 0$. This means that if voter classes are homogeneous w.r.t the average node degree, then such a correlation will not emerge in the pair approximation system. Hence, from Eq. \eqref{eqn:ave:degree:quenched} we get:
\begin{eqnarray}
    \sum_{s'}\sum_{\eta'}e_{ss'}^{\eta\eta'} = c_{s,\eta}
\end{eqnarray}
which reduces the number of independent variables by $2S$ for the quenched system and $S$ for the annealed one, allowing us to describe the evolution of the system solely with edge concentrations.

\section{Results}
\begin{figure}[htbp]
    \centering
    \includegraphics[width=\linewidth]{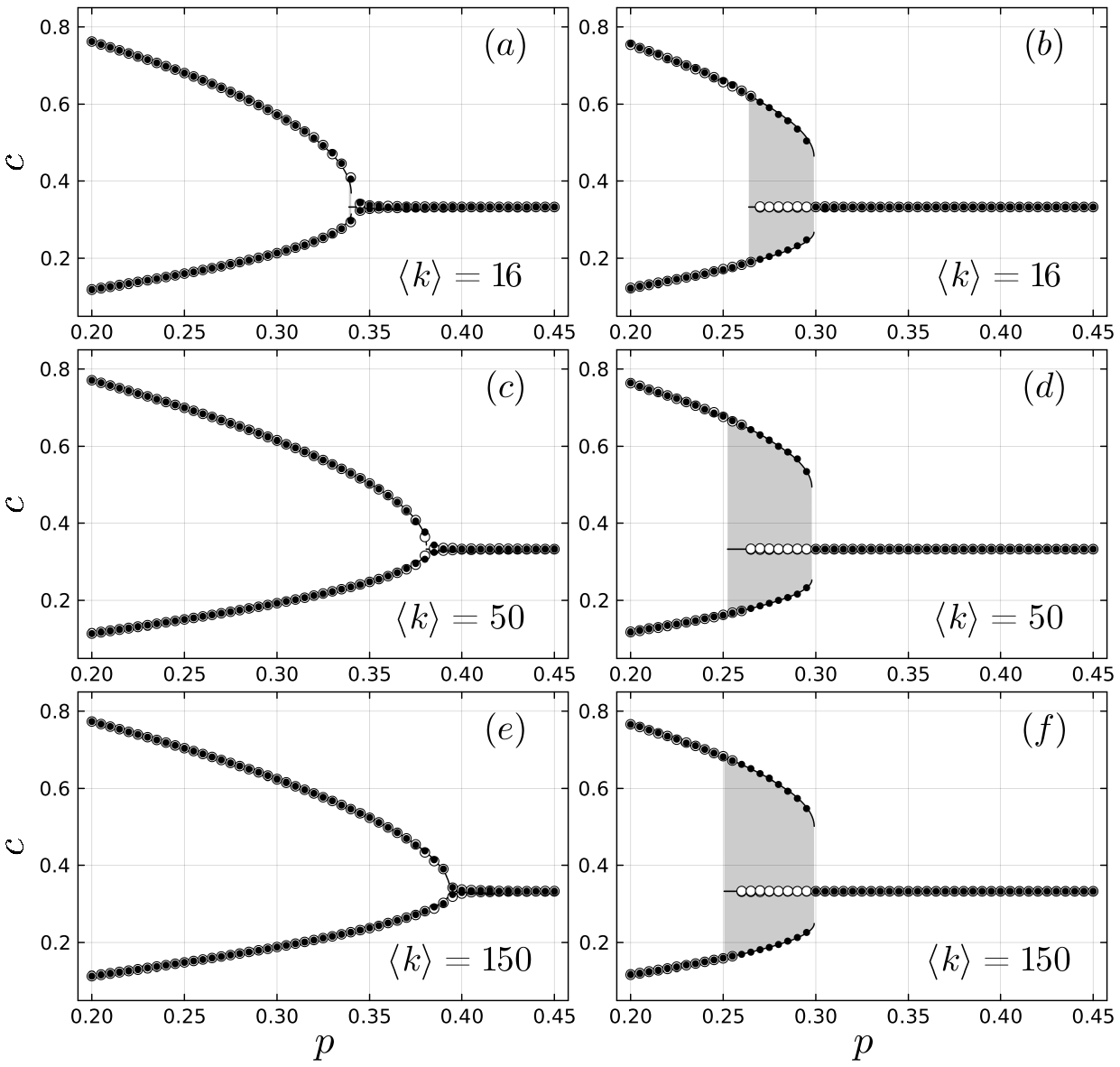}   
    \caption{\textbf{The influence of graph density on phase transitions within PA and simulations.} Stationary values of state concentrations (of the highest and lowest occupied state) obtained from PA (solid lines) and Monte Carlo simulations on \textbf{random graphs} (markers) for the three-state ($S=3$) $q$-voter model with $q=2$ in the annealed (left panels: a, c, e) and quenched (right panels: b, d, f) approaches. Empty symbols correspond to results obtained from an initially disordered state ($1/3$ of voters in each state at $t=0$), whereas filled symbols correspond to the initial condition where all voters are in the same state. Panels (a)-(b) show results for $\langle k \rangle = 16$, (c)-(d) for $\langle k \rangle = 50$, and (e)-(f) for $\langle k \rangle = 150$. The shaded area highlights the hysteresis region obtained from PA.}
    \label{fig:q2_rr}
\end{figure}

\begin{figure}[!htbp]
    \centering
    \includegraphics[width=\linewidth]{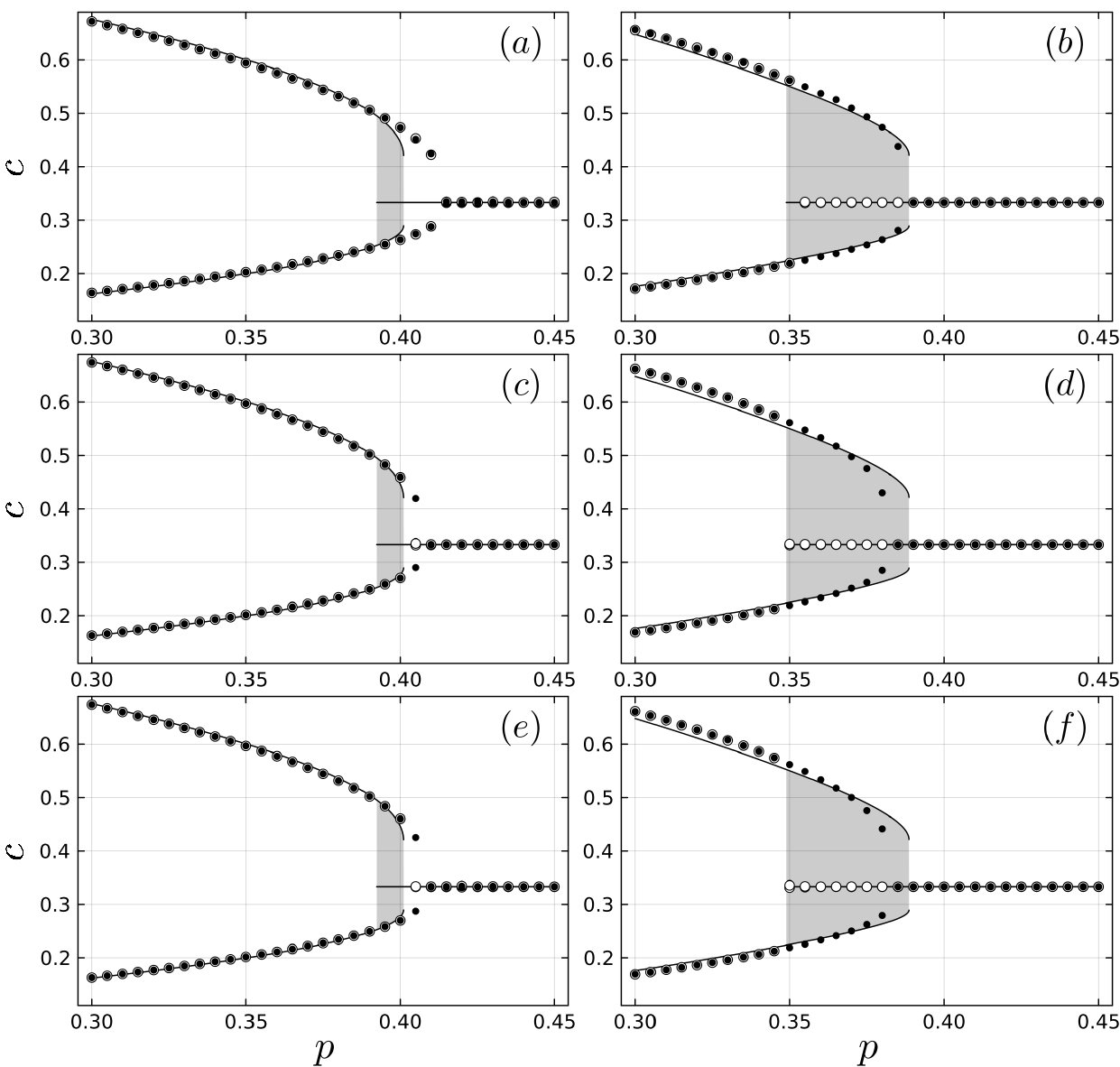}
    \caption{\textbf{The influence of graph type with a fixed density of $\langle k\rangle = 16$ on phase transitions within PA and simulations.} Stationary values of state concentrations (of the highest and lowest occupied state) obtained from PA (solid lines) and Monte Carlo simulations (markers) for the three-state $q$-voter model with $q=3$ and $\langle k\rangle = 16$ on \textbf{Barabási–Albert} (top panels: a, b), \textbf{random regular} (middle panels: c, d) and \textbf{random} (bottom panels: e, f) graphs in the annealed (left: a, c, e) and quenched (right: b, d, f) disorder. Empty shapes mark results obtained from initial disorder, i.e. $1/3$ of voters in each state at $t=0$, while filled shapes correspond to the initial condition in which all voters are in the same state. Shaded area highlights the hysteresis region obtained from PA.}
    \label{fig:q3}
\end{figure}

The aim of this work was to check whether quenched anticonformity can induce discrete phase transitions and hysteresis not only for the complete graph ($\langle k \rangle = N-1$) but also for sparser networks. In addition to the analytical PA results, we also conducted computer simulations on random graphs, random regular graphs, and Barabási–Albert scale-free networks~\cite{Barabasi1999} to verify whether discontinuous transitions actually occur. This is crucial because PA can predict discontinuous phase transitions and hysteresis even in cases where simulations suggest continuous transitions \cite{Abramiuk-Szurlej2021DiscontinuousGraphs,Jedrzejewski2022PairNetworks}.
To generate random graphs we use a Watts-Strogatz algorithm~\cite{Watts1998} with $\beta = 1$, for which every edge in the graph is randomly rewired and every vertex has at least $\langle k \rangle/2$ neighbors. We present results for large graphs, i.e. $N = 10^6$ for $q=2$ and $N=5\cdot10^6$ for $q=3$, as for graphs of this size the outcomes of simulations and PA closely agree, as shown in Figs. \ref{fig:q2_rr} and \ref{fig:q3}. 

We focus on $\langle k \rangle << N$, with values inspired by empirical research showing that human social networks naturally organize into fractal layers comprising groups of sizes 1.5, 5, 15, 50, 150, 500, 1500 and 5000, observed in both face-to-face and digital interactions; for review see \cite{Dunbar2021FriendsRelationships,Dunbar2024TheOn}. The smallest layers 1.5 and 5 correspond to our most intimate relationships, such as a romantic partner, close family, or best friends, while layers 15 and 50 represent good friends and wider circles of casual but trusted companions, and the 150, known as Dunbar's Number, corresponds to our core community, the people we would invite to important life events.

In this work, we consider networks with the average node degree of 16, 50, and 150. The adjustment from 15 to 16 neighbors is made purely for computational reasons, since some of the graph generating algorithms that we consider produce even average node degree. Sparser networks are difficult to study due to model constraints, particularly for larger $q$, because reliable results require $\langle k \rangle$ to be sufficiently large relative to $q$ \cite{Abramiuk-Szurlej2021DiscontinuousGraphs,Ramirez2024OrderingModels}. For the same reason, we limit our attention to $q = 2$ (Fig. \ref{fig:q2_rr}) and $q=3$ (Fig. \ref{fig:q3}). These values correspond to interacting groups of 3 and 4 people (one target agent and $q$ source agents), which were recently used in social experiments on strategic anticonformity \cite{Dvorak2024StrategicReward}. Moreover, we focus on three-state systems, $S = 3$, for which the puzzling effect of switching from a continuous to a discontinuous phase transition under quenched anticonformity was observed on a complete graph \cite{Nowak2022SwitchingDisorder}. 

\begin{figure}[!htbp]
    \centering
    \includegraphics[width=0.8\linewidth]{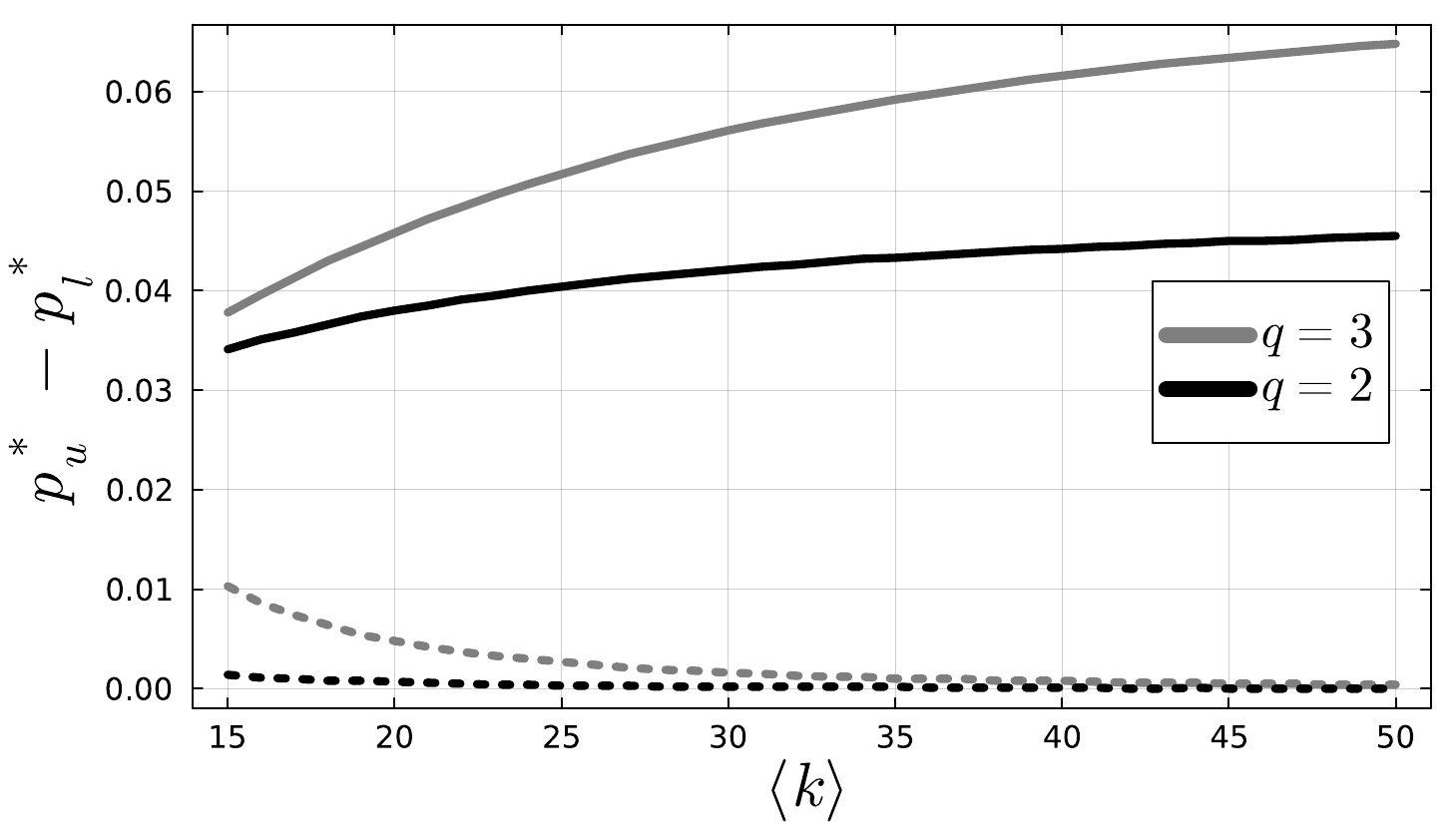} 
    \caption{\textbf{The influence of graph density on the width of hysteresis within PA.} The width of hysteresis, calculated as  the difference between the upper and lower spinodals and represented by the shaded areas in Figs. \ref{fig:q2_rr} and \ref{fig:q3} with respect to the average node degree $\langle k \rangle$ obtained from PA for the three-state $q$-voter model with the annealed (dotted lines) and quenched (solid lines) disorder.}
    \label{fig:hysteresis}
\end{figure}

Results shown in Figs. \ref{fig:q2_rr} and \ref{fig:q3} demonstrate that, under quenched disorder, a discontinuous phase transition indeed occurs. In the quenched approach, both hysteresis and the jump in stationary state concentrations increase with increasing $\langle k \rangle$, while in the annealed approach the situation is somewhat reversed: the transition becomes sharper for smaller values of $\langle k \rangle$. The corresponding increase in hysteresis for the quenched case and decrease for the annealed case with respect to the average node degree $\langle k \rangle$, is clearly shown in Fig. \ref{fig:hysteresis}. 

We determine the width of hysteresis, presented in Fig. \ref{fig:hysteresis}, as the difference between the upper spinodal (the greatest value of $p$ for which there exists an ordered stationary solution) and the lower spinodal (the lowest value of $p$ for which the disordered state is a stationary solution). For the annealed disorder, the results of PA indicate a non-zero width of hysteresis for small values of $\langle k \rangle$, which is clearly visible for $q=3$ (left panels of Fig. \ref{fig:q3}), and much less so for $q=2$ (top left panel of Fig. \ref{fig:q2_rr}). 
However, hysteresis increases as $q$ increases and $\langle k \rangle$ decreases, and PA is known to yield inaccurate approximations in these cases \cite{Abramiuk-Szurlej2021DiscontinuousGraphs,Ramirez2024OrderingModels}. In contrast, for the quenched disorder the width of hysteresis grows with the average node degree. When $\langle k \rangle$ approaches the complete-graph limit, the PA predictions for both annealed and quenched scenario are in agreement with the mean-field and simulation results~\cite{Nowak2022SwitchingDisorder}. For sparser networks, the pair approximation agrees with simulations only in the quenched case, where hysteresis is clearly observed, as shown in the right panels of Figs. \ref{fig:q2_rr} and \ref{fig:q3}. In the annealed case, the hysteresis observed in simulations is absent or miniscule (with faint traces visible in panels (c) and (e) of Fig. \ref{fig:q3}) and remains inconsistent with PA predictions.

\FloatBarrier
\section{Conclusions}
The study of the role of quenched disorder has a long tradition in the physics of phase transitions, and it is generally known that such disorder rounds or even completely eliminates discontinuous phase transitions \cite{Imry1975Random-fieldSymmetry,Binder1987TheoryTransitions,Aizenman1989RoundingDisorder,Vojta2014CriticalityRegions}. In the field of opinion dynamics, the problem has been investigated both in terms of network connections \cite{Baron2021PersistentDisorder,Mihara2023CriticalModel} and in terms of agent behavior \cite{Nowak2022SwitchingDisorder,Jedrzejewski2022PairNetworks}. The latter aspect is related to the so-called person-situation debate \cite{Sznajd-Weron2014IsVice-versa}, a long-standing discussion in psychology about whether behavior is determined primarily by stable personal traits (the “person” view, naturally corresponding to quenched disorder) or by external situational factors (the “situation” view, corresponding to annealed disorder). More recently, annealed and quenched dynamics have been compared in the mean-field (well-mixed population) limit within a general framework for binary-choice dynamics, in which agents update their states using two mechanisms such as conformity and anticonformity, among others. Within this framework, the conditions on transition probabilities under which annealed and quenched dynamics become equivalent were identified \cite{Jedrzejewski2025EveryoneDiffusion}.

Comparisons between quenched and annealed approaches have also been carried out on various random graphs, but only within specific models: (1) the binary $q$-voter model with independence and (2) the binary $q$-voter model with anticonformity. For the multistate $q$-voter model, such a comparison has so far been performed only on the complete graph: for the model with independence \cite{Nowak2021DiscontinuousDisorder} and for the model with anticonformity \cite{Nowak2022SwitchingDisorder}. An intriguing result has been obtained for the latter. On the complete graph, which corresponds to the mean-field limit, the annealed version of the $q$-voter model with anticonformity displays only continuous phase transitions, regardless of the number of states $S$ and the influence group size $q$. In contrast, the quenched version exhibits discontinuous phase transitions for $S \geq 3$ and $q \geq 2$. The aim of this work was to test whether this unexpected effect also persists on sparser graphs.

Our results show that quenched anticonformity can indeed induce discontinuous phase transitions and hysteresis not only on the complete graph but also on sparser networks. We verified this effect using pair approximation and Monte Carlo simulations on random graphs, random regular graphs, and Barabási-Albert scale-free networks. In all these cases, quenched dynamics consistently produce discontinuous transitions, with the hysteresis width increasing with network density. In contrast, under annealed dynamics, continuous transitions sharpen as the average degree decreases, leading to discontinuous transitions in PA predictions but only minimal or no hysteresis in MC simulations. Such inconsistencies between PA and simulation results have also been reported for several versions of the binary $q$-voter model with anticonformity \cite{Abramiuk-Szurlej2021DiscontinuousGraphs,Jedrzejewski2022PairNetworks}.

Empirical evidence shows that anticonformity can indeed emerge and be induced in social groups, demonstrating that it is not merely a theoretical construct \cite{Dvorak2024StrategicReward}. Theoretically, it has been shown that anticonformity can depolarize already polarized groups \cite{Lipiecki2025DepolarizingAnticonformity}, a surprising and nontrivial effect. Together, these findings suggest that anticonformity can act as a highly intriguing social response, puzzlingly shaping collective opinion dynamics and influencing phase transitions in social systems. Future research should examine whether analogous phenomena occur in other multistate opinion dynamics models, shedding light on the general role of anticonformity in shaping collective behavior.

\section*{Author contributions}
A.L. was responsible for all analytical calculations and Monte Carlo simulations. K. S-W. was responsible for supervising the research and funding acquisition. Both authors were writing, reviewing, and editing the manuscript.

\section*{Acknowledgments}
This work was partially supported by funds from the National Science Centre (NCN,Poland) through grants no. 2019/35/B/HS6/02530 (to AL) and 2023/51/I/HS6/02269 (to KSW). 
`

\end{document}